\pdfoutput=1
\documentclass[11pt]{article}
\usepackage{jheppub}

\usepackage{amsmath,amsfonts,amssymb,amsthm,amstext,amscd,eucal}
\usepackage{graphicx}

\usepackage{color}

\newcommand{\bZ}{\mathbb{Z}}
\newcommand{\bC}{\mathbb{C}}

\newcommand{\bP}{\mathbb{P}}

\newcommand{\cC}{\mathcal{C}}
\newcommand{\cE}{\mathcal{E}}
\newcommand{\cF}{\mathcal{F}}

\newcommand{\cO}{\mathcal{O}}
\newcommand{\cS}{\mathcal{S}}

\newcommand{\ov}{\overline}

\newcommand{\sO}{\mathsf{O}}

\usepackage{ifpdf}
\usepackage[boxsize=0.5em,aligntableaux=center]{ytableau}

\newcommand{\symm}{\ydiagram{2}}
\newcommand{\asymm}{\ydiagram{1,1}}
\newcommand{\fund}{\ydiagram{1}}

\DeclareMathOperator{\Vol}{Vol}

\DeclareMathOperator{\Ext}{Ext}
\renewcommand{\Re}{\mathrm{Re}\,}

\title{D-brane instantons on non-Spin cycles}

\author[a]{Per Berglund}
\author[b]{and I\~naki Garc\'ia-Etxebarria}

\affiliation[a]{Department of Physics, University of New Hampshire,
  Durham, NH 03824, USA}
\affiliation[b]{Theory Group, Physics
  Department, CERN, CH-1211, Geneva 23, Switzerland}

\emailAdd{per.berglund@unh.edu}
\emailAdd{inaki@cern.ch}

\abstract{We show that non-Spin cycles in type IIB Calabi-Yau
  orientifold compactifications with vanishing $B$ field naturally
  support orientifold-invariant D-brane instantons. These instantons,
  associated to the holomorphic cotangent bundle of the non-Spin
  cycle, can lead to non-perturbative corrections to the
  superpotential.}

\begin{document}

\maketitle

\section{Introduction}

Moduli stabilization is a fundamental problem when constructing
realistic models of particle physics in string theory. In IIB string
theory (or F-theory), the closed string moduli in the K\"ahler sector
are stabilized by non-perturbative contributions to the
superpotential. There are various promising scenarios in
which non-perturbative effects, in the form of D-brane instantons,
give rise to moduli stabilization, see e.g.
\cite{Denef:2004dm,Balasubramanian:2005zx}.

When studying contributions to the superpotential, a natural class of
candidate instantons arises from D-branes wrapping rigid cycles. These
cycles are often non-Spin, and in particular del Pezzo surfaces occur
frequently. For non-Spin cycles one needs to consider a Spin$^c$
structure for the worldvolume degrees of freedom instead of an
ordinary Spin structure \cite{Freed:1999vc}.

In the case of ordinary line bundles this is commonly phrased as
introducing a ``half-quantized bundle'' on the instanton. However,
this half-quantized bundle necessarily clashes with the requirement of
invariance under the orientifold involution. It is well know how to
restore invariance under the orientifold by introducing a
half-quantized $B$-field (see for example
\cite{Blumenhagen:2012kz,Cicoli:2012vw}).

Nevertheless, whether there is a non-perturbative contribution in the
original $B=0$ case remains an open question. In this paper we show
that when $B=0$ there does exist an invariant rank 2 instanton with
the right properties to contribute to the superpotential.

\section{Invariant instantons and Spin$^c$ structures}

We will use the sheaf description of D-branes, see
\cite{Sharpe:2003dr} for a good review.  In our case these sheaves
will be ordinary vector bundles with support on the instanton
divisor. Consider the case of a surface $\cS$ in an ambient Calabi-Yau
space $X$, with inclusion map $i:\cS\hookrightarrow X$, and take a
sheaf $i_*\cE$ describing the instanton wrapping $\cS$. This
corresponds to an instanton with bundle $\cE\otimes
\sqrt{K_{\cS}^\vee}$ \cite{Katz:2002gh,Sharpe:2003dr}. In particular,
this construction automatically gives rise to a well-defined Spin$^c$
structure on the D-brane. A necessary condition, when $B=0$, for the
instanton to be invariant under the orientifold involution is
\cite{Diaconescu:2006id}:
\begin{align}
  \label{eq:invariance-condition}
  \sigma^*(\cE^\vee) \otimes K_\cS = \cE\, ,
\end{align}
where $\sigma$ is the orientifold involution. A solution to this
equation is given by $\cE=\Omega$, where $\Omega$ is the dual of the
holomorphic tangent bundle of $\cS$, or equivalently the sheaf of
holomorphic one-forms. In the remainder of this paper we show in a
particular example that the corresponding instanton has the right
structure to contribute to the non-perturbative superpotential.

\section{An invariant instanton on $\bP^2$}

For concreteness we focus on (the complex Calabi-Yau cone over)
$\bP^2$, the simplest example of a non-Spin manifold, and return to
the general case in the conclusions.  We parameterize $X$, our
Calabi-Yau manifold, by the coordinates $(s,t,u,w)$, identified under
the rescalings $(s,t,u,w)\sim (\lambda s, \lambda t, \lambda u,
\lambda^{-3} w)$, for all $\lambda \in \bC^*$. The $\bP^2$ wrapped by
the instanton is located at $z=0$, and the orientifold involution that
fixes the $O7^-$ plane acts geometrically as
\begin{align}
  \label{eq:orientifold}
\sigma:\,s\leftrightarrow -s\, .
\end{align}

Since we are interested in contributions to the superpotential, our
task is to show that the instanton only has the two universal zero
modes $\theta^{\alpha}$, with every other possible zero mode absent,
see \cite{Blumenhagen:2009qh} for a review of D-brane instanton
physics. These zero modes can be either neutral $\ov\tau_{\dot\alpha}$
modes, coming from the instanton being non-invariant, neutral
deformation modes coming from the instanton not being rigid, or
charged zero modes coming from massless strings between the instanton
and the background D-branes.

\subsection{Neutral zero modes}

The absence of neutral zero modes (i.e., those of deformation and
$\ov\tau_{\dot\alpha}$ types) can be rephrased as the instanton being
rigid and of $O(1)$ type, and can be shown as follows. First of all,
since $\bP^2$ is rigid, there are no deformation modes. By explicitly
writing the transition functions it is not difficult to see that
$\Omega$ is invariant under the orientifold involution
\eqref{eq:orientifold}, i.e. it
satisfies~\eqref{eq:invariance-condition}. In order to prove that the
instanton is of $O(1)$ type we need to show that the gauge symmetry on
the instanton in the absence of the orientifold is $U(1)$, and that
the orientifold projects this to the $O(1)$ component (as opposed to
projecting down to $U\!Sp$). The fact that the rank 2 bundle preserves
a $U(1)$ gauge group can be understood by noting that the structure
group of $\bP^2$, being K\"ahler, is $U(2)$, while the local symmetry
on the rank 2 bundle is also $U(2)$. The preserved gauge group is the
commutant of both factors, and it is thus $U(1)$. A more formal but
also more systematic way of showing this comes from simply computing
$\Ext^p(i_*\Omega,i_*\Omega)$, which is the space of massless adjoint
fermions \cite{Aspinwall:2004jr}. For this particular bundle we have
$\Ext^0(i_*\Omega,i_*\Omega)=\bC$, and by Serre duality on $X$ we also
have $\Ext^3(i_*\Omega,i_*\Omega)=\bC$. The $\Ext$ groups can be shown
to vanish for all other values of $p$. The fact that the projection
induced by the orientifold involution is of type $O(1)$ can be seen by
a local argument of monodromy around the orientifold locus
\cite{Cvetic:2009ah}, which shows that the $\ov\tau_{\dot\alpha}$
modes are the ones projected out. A different way of understanding
these last two facts will be given below when we study the instanton
at small volume.

\subsection{Charged zero modes}
\label{sec:charged-zero-modes}

We are left to show that the charged zero modes --- massless strings
stretching between the instanton and background D-branes --- are
absent. In our particular configuration the latter correspond to the
$D7$ branes required to cancel the tadpole induced on the hyperplane
of the $\bP^2$ by the $O7^-$. We perform the zero mode computation at
a particular point in the moduli space of the $D7$ branes where they
restrict to a stack of 4 D7 branes wrapping a quadratic curve on
$\bP^2$. If there is a non-vanishing superpotential at this particular
point in moduli space, by continuity of the superpotential this shows
that generically the superpotential is non-vanishing.

The spectrum of charged zero modes in our particular configuration is
counted by \cite{Katz:2002gh}:
\begin{align}
  \label{eq:complete-spectrum}
  \Ext^p(j_*\cF, i_*\Omega) = H^{p-1}(\cC,
  \Omega|_\cC\otimes\cF^\vee|_\cC\otimes N_{\cC|\bP^2})\, .
\end{align}
Here $\cC$ denotes the curve (a quadratic $\bP^1\subset\bP^2$) where
the $D7$ branes and the instanton worldvolume intersect,
$N_{\cC|\bP^2}$ is the normal bundle of this $\bP^1$ in $\bP^2$, given
by $\cO(4)$ and $\cF$ is the bundle of the $D7$ flavor brane stack,
with $j$ the embedding of the D7 stack into $X$. (We will only need
the restriction of $j$ to $\cC$, given in eq.~\eqref{eq:j-map} below.)
In order to calculate the spectrum~\eqref{eq:complete-spectrum} we
thus need to compute the restrictions of the appropriate cotangent
bundles to $\cC$. Since $\cC$ is just a $\bP^1$, by a well-known
result due to Grothendieck we have that $\Omega$, being of rank 2,
splits into a sum of two line bundles when restricted to $\cC$:
$\Omega|_\cC=\cO(m)\oplus \cO(n)$. In addition, the sums of the
individual degrees of the line bundles are constrained by the
following relation:
\begin{align}
  n+m = \int_\cC c_1(\Omega|_\cC) = \int_{\bP^2} c_1(\Omega) \wedge
  2\ell = -6\, .
\end{align}

To determine $m$ and $n$ separately we follow the approach in
\cite{KatzFiniteness}.  Let us focus on the local neighborhood of
$\bP^2$, the instanton cycle. We take $\cC$, the quadratic curve in
$\bP^2$ wrapped by the $D7$ branes, to be given by $s^2 - tu=0$. We
can also parameterize $\cC$ by the projective coordinates $(x_1,x_2)$
with an embedding in $X$ given by
\begin{align}
  \label{eq:j-map}
  (x_1,x_2)\mapsto (s,t,u,w)=(x_1x_2, x_1^2, x_2^2,0). 
\end{align}

Let us first compute the
restriction $T_X|_\cC$ of the tangent bundle of the Calabi-Yau $X$ to
$\cC$, using the toric Euler sequence restricted to $\cC$:
\begin{align}
  0 \to \cO_\cC \xrightarrow{i} (\cO_X(1)^{\oplus 3} \oplus
  \cO_X(-3))|_\cC \xrightarrow{f} T_X|_\cC \to 0 \, .
\end{align}
It is easy to see that $\cO_X(n)|_\cC=\cO_\cC(2n)$, for instance by
computing the following integral:
\begin{align}
  \begin{split}
    k  & = \int_\cC c_1(\cO_\cC(k)) = \int_\cC c_1(\cO_X(n)|_\cC) \\
    & = \int_{\bP^2} n\ell \wedge 2\ell = 2n\, .
  \end{split}
\end{align}
The exact sequence to study is thus:
\begin{align}
  \label{eq:restricted-Euler}
  0 \to \cO_\cC \xrightarrow{i} \cO_\cC(2)^{\oplus 3} \oplus
  \cO_\cC(-6) \xrightarrow{f} T_X|_\cC \to 0 \, .
\end{align}
The map $i$ is given by the inclusion of $\cC$ into $X$, which acts on
the sections as:
\begin{align}
  i = \begin{pmatrix}
    x_1x_2 \\ x_1^2 \\ x_2^2 \\ 0
  \end{pmatrix}\, .
\end{align}
The map $f$ is defined by imposing exactness
of~\eqref{eq:restricted-Euler}, i.e., $f\circ i={}^t\!(0\,\,0\,\,0)$,
and it is thus given by:
\begin{align}
  f = \begin{pmatrix}
    x_1 & -x_2 & 0 & 0\\
    x_2 & 0 & -x_1 & 0\\
    0 & 0 & 0 & 1
  \end{pmatrix}\, .
\end{align}
Imposing that the degrees of the bundles in
\eqref{eq:restricted-Euler} make sense, we thus obtain
$T_X|_\cC=\cO_\cC(3)\oplus \cO_\cC(3)\oplus \cO_\cC(-6)$.

In order to find the restriction of $T_\cS$ from the restriction of
$T_X$ we use the adjunction formula:
\begin{align}
  0 \to T_D|_\cC \xrightarrow{K} T_X|_\cC \xrightarrow{L}
  \cO_X(D)|_\cC \to 0\, ,
\end{align}
where $D=\bP^2$ is the divisor of interest. In our particular case
this becomes
\begin{align}
  \label{eq:adjunction}
  0 \to T_D|_\cC \xrightarrow{K} \cO_\cC(3)\oplus\, \cO_\cC(3)\oplus
  \cO_\cC(-6) \xrightarrow{L} \cO_X(D)|_\cC \to 0\, ,
\end{align}
with $D$ given by $w=0$. In general, if $D$ is defined by a
homogeneous polynomial equation $P=0$, we have that the map (in the
exact sequence before restriction) $\widehat L\colon T_X\to \cO(D)$ is
given by $\sum p_i\frac{\partial}{\partial t_i}\to \sum
p_i\frac{\partial P}{\partial t_i}$, where $t_i$ are the toric
coordinates $(s,t,u,w)$, and $p_i$ a homogeneous polynomial of the
same degree as $t_i$. Let us perform explicitly the restriction of the
tangent bundle to $\bP^2$.  In this case $P=w$. The map
$L=(a_1\,\,a_2\,\,a_3)$ is then given by the solution to:
\begin{align}
  (a_1\,\,a_2\,\,a_3) \begin{pmatrix}
    s & -t & 0 & 0\\
    t & 0 & -s & 0\\
    0 & 0 & 0 & 1
  \end{pmatrix} = \begin{pmatrix} 0 & 0 & 0 & 1
    \end{pmatrix}\, .
\end{align}
Solving these equations, we have $(a_1\,\,a_2\,\,a_3)=(0\,\,0\,\,1)$,
which is compatible with the fact that $O_X(D)|_\cC=\cO_\cC(-6)$. This
tells us that the map $L$ is simply picking the last component in the
$\cO_\cC(3)\oplus \cO_\cC(3)\oplus \cO_\cC(-6)$ bundle, discarding the
rest. From exactness of~\eqref{eq:adjunction}, this implies that
$T_{\bP^2}|_\cC = \cO_\cC(3)\oplus \cO_\cC(3)$, or equivalently by
dualization $\Omega|_\cC = \cO_\cC(-3)\oplus \cO_\cC(-3)$.

We also need to find the restriction to $\cC$ of the bundle $\cF$ on
the $D7$ flavor brane stack. The situation here is less well defined,
as in principle in order to specify the bundle one needs to consider
the whole compact divisor wrapped by the flavor brane. However, at
least locally it makes sense to choose a trivial bundle. In
particular, the total space $\sO$ of the $\cO(-3)$ fiber over the
quadratic $\bP^1$ curve $\cC$ has the structure of a toric space with
coordinates $(x_1,x_2,p)$, identified under the rescaling
$(x_1,x_2,p)\sim (\lambda x_1, \lambda x_2, \lambda^{-6} p)$, with
$\lambda\in\bC^*$. The local intersections are given by $x_1\cdot
p=x_2\cdot p = 1$, $p\cdot p=-6$. We identify $p=0$ with $\cC$. The
canonical class of the space is then given by $K_\sO=-D_1-D_2 - D_p$,
and we immediately see that locally the cycle wrapped by the flavor
brane is Spin, since the intersection of $K_\sO$ with the local class
$D_p$ is even. So, locally we can choose a trivial (physical) flux on
the flavor brane. In the conventions used in this paper, where we work
in terms of the sheaves defining the D-branes, we have that
$\cF=\sqrt{K_\sO}$ (so $\cF\otimes \sqrt{K_\sO^\vee}$ is trivial). In
particular, we have the restriction
$\cF|_\cC=\sqrt{K_\sO|_\cC}=\cO_\cC(2)$, since
$K_\sO|_\cC=\cO_\cC(4)$. As one may expect, this choice is also
invariant under the orientifold action \cite{Diaconescu:2006id}:
\begin{align}
  \begin{split}
    \cF|_\cC = \cO_\cC(2) \longrightarrow & \,\, \cF^\vee|_\cC
    \otimes
    K_\sO|_\cC \\
    & = \cO_\cC(-2)\otimes \cO_\cC(4) = \cO_\cC(2)
  \end{split}
\end{align}
where we have again restricted to the curve, in order to not  have to
involve global information.

Now that we have obtained the restriction of the gauge bundles to the curves,
we can easily compute the full spectrum by plugging $\cF|_\cC$ and
$\Omega|_\cC$ into~\eqref{eq:complete-spectrum}:
\begin{align}
  \begin{split}
    \Ext^1(j_*\cF, i_*\cE) & = H^0(\cC,
    \cO_\cC(-3)^{\oplus 2}\otimes \cO_\cC(-2) \otimes
    \cO_\cC(4))\\
    & = H^0(\cC, \cO_\cC(-1)^{\oplus 2}) = 0\, .
  \end{split}
\end{align}
Similarly, using Serre duality and the fact that $X$ is Calabi-Yau we
have that $\Ext^1(i_*\cE, j_*\cF)=\Ext^2(j_*\cF, i_*\cE)=H^1(\cC,
\cO_\cC(-1)^{\oplus 4})=0$. So, there are no charged zero modes, and
we see that the instanton does give rise to a non-vanishing
superpotential $W_{\rm non-pert}=A_{1-{\rm loop}}\,e^{-S_{inst}}$,
where $A_{1-{\rm loop}}$ is the non-zero one-loop determinant and
$S_{inst}$ is the instanton action. In general we have that
$\Re(S_{inst})=|Z(i_*\Omega)|/g_s$, with $Z(i_*\Omega)$ the central
charge of the instanton. At large volume this reduces to
$\Re(S_{inst})=2\Vol(D)/g_s$, with $\Vol(D)$ the volume of the $\bP^2$
on which we are wrapping the instanton. Notice in particular the
factor of 2 multiplying the volume, coming from the fact that we are
considering a rank two stack. So at large volume there is an extra
suppression of the non-perturbative effect compared to the case of a
hypothetical rank one instanton on the same cycle which --- had there
been such a contribution to the superpotential --- would have gone as
$\Re(S_{inst})=\Vol(D)/g_s$.\footnote{We would like to thank the
  referee for pointing out a missing factor of two in the expression
  for the instanton action given in the original version of this
  paper, and emphasizing the resulting suppression.}  As we go towards
small volumes $\alpha'$ corrections to the large volume expression for
$Z(i_*\Omega)$ become important, and the suppression due to the rank
of the bundle is less pronounced. (A couple of particularly
interesting points in K\"ahler moduli space are the quiver point
analyzed in the next section, in which $Z(i_*\Omega)$ coincides with
the central charge for certain rank one branes wrapping the same
cycle, and deep in the orbifold phase, where one can even have
$Z(i_*\Omega)=0$ \cite{Aspinwall:2004jr}.)

Note that the instanton contribution to the superpotential may vanish
if $\cF$ is another vector bundle on the $D7$ flavor brane stack. For
example, a different natural choice is $\cF=\Omega_\sO$, the cotangent
bundle for the divisor $\sO$, with $\cF$ invariant under the
orientifold action and the Freed-Witten anomaly cancellation
conditions satisfied. It is a simple exercise using the technology
described above to show that $\Omega_\sO|_\cC=\cO_\cC(-2)\oplus
\cO_\cC(6)$. Computing the zero modes in this case we obtain:
\begin{align}
  \begin{split}
    \Ext^1(j_*\cF, i_*\cE) & = H^0(\cC,
    \cO_\cC(-3)^{\oplus 2}\otimes\\
    &
     \phantom{= H^0(} (\cO_\cC(2)\oplus\cO_\cC(-6))\otimes
    \cO_\cC(4))\\
    & = H^0(\cC, \cO_\cC(3)^{\oplus 2}\oplus
    \cO_\cC(-5)^{\oplus 2})  \\
    & = \bC^8\, ,
  \end{split}
\end{align}
and similarly $\Ext^1(i_*\cE, j_*\cF)=\bC^8$. Thus, there are eight
pairs of vector-like zero modes, and the contribution to the
superpotential vanishes. There is no contradiction between this result
and the one above, since both bundles live in different components of
the moduli space. In particular, the two bundles induce different
amounts of $D3$ charge, and thus are connected by emission/absorption
of mobile $D3$ branes. Thus, this provides partial information about the
dependence of the instanton action on the moduli of (mobile) $D3$
branes. In particular, it shows the existence of zeros in the one-loop
determinant of the instanton contribution to the superpotential when
the $D3$ branes hit the instanton~\cite{Ganor:1996pe}.

\section{Small volume interpretation}

With the contribution of the instanton described by $i_*\Omega$ to the
superpotential established, let us present an alternative viewpoint
that arises when we take the configuration to the quiver point.
Specifically, let the volume of the $\bP^2$ be zero, which turns the
smooth space $X$ into the well-known $\bC^3/\bZ_3$ orbifold.

We also momentarily forget about the instanton, and instead consider
$2N$ regular $D3$ branes probing the singularity of the $\bC^3/\bZ_3$
orbifold, in addition to the $O7^-/D7$ stack.  Since we are setting
$B=0$ this is the quiver point for the $D3$ branes
\cite{Aspinwall:2004jr}, and the $D3$ branes decompose into fractional
branes as they hit the singularity. The resulting spectrum is obtained
using CFT methods, or more generally dimer model methods
\cite{Franco:2007ii}. One obtains a theory with $U\!Sp(2N)\times
U(2N)$ gauge group and $U(4)$ flavor group, with the following matter
content:
\begin{align}
  \label{eq:orientifolded-quiver}
  \begin{array}{l|cc|c}
    & U\!Sp(2N) & U(2N) & [U(4)]\\
    \hline
    X^i\quad (i\in 1,2,3) & \fund & \ov\fund & {\bf 1}\\
    A^j\quad (j\in 1,2) & {\bf 1} & \asymm & {\bf 1}\\
    S & {\bf 1} & \symm & {\bf 1}\\
    Q & {\bf 1} & \fund & {\bf 4}
  \end{array}
\end{align}

The same result can also be derived using large volume language. The
three elementary fractional branes can be described by the objects
$i_*\cO(-1)[0]$, $i_*\Omega[1]$ and $i_*\cO(-2)[2]$. (Here the numbers
in square brackets denote the position of the given sheaf in the
associated sheaf complex describing the brane \cite{Aspinwall:2004jr},
and will not be essential in what follows.) The main point is that the
orientifold action \eqref{eq:orientifold} leaves the $i_*\Omega[1]$
brane invariant, leading to the $U\!Sp(2N)$ factor, while it exchanges
$i_*\cO(-1)[0]$ and $i_*\cO(-2)[2]$, giving rise to the $U(2N)$ factor
(once we take $2N$ such objects).  The matter content can be
determined by computing $\Ext$ groups between the fractional branes,
and the $U(4)$ global symmetry factor comes from the $D7$ brane stack
wrapped on the quadratic divisor.

The most important part of this discussion for our purposes is the
gauge group $U\!Sp(2N)$ associated to the invariant node. Gauge
instantons for $U\!Sp(2N)$ field theories have gauge symmetry $O(1)$.
In particular, the euclidean $D3$ brane describing the field theory
instanton will be of type $O(1)$, i.e., given by the same sheaf as the
field theory brane. Since the $U\!Sp$ stack is associated with
$i_*\Omega$ the gauge instanton in this node is precisely the
instanton we have been studying. In fact, it is not necessary to have
$N\neq 0$. If $N=0$ the invariant node gives a $USp(0)$
``gauge group'', which also gives rise to non-perturbative string
theory dynamics
\cite{Aganagic:2003xq,Intriligator:2003xs,Aganagic:2007py}, due to
$O(1)$ D-brane instantons \cite{GarciaEtxebarria:2008iw}.

\section{Conclusions}

We have shown the existence of an $O(1)$ type instanton $i_*\Omega$,
invariant under the orientifold involution, in the particular case of
the complex Calabi-Yau cone over $\bP^2$.  Although the discussion
focused on a particular example, it is clear that the analysis holds
for essentially any rigid cycle, and that the basic phenomenon will be
ubiquitous in type IIB string compactifications.

A natural extension of this work would be to consider higher rank
instantons, ideally classifying all bundles
satisfying~\eqref{eq:invariance-condition}.  Furthermore, even
rigidity is not a necessary condition for instanton contributions when
worldvolume fluxes are taken into account~\cite{Bianchi:2011qh}.  It
would be interesting to study the general conditions under which
non-rigid, non-Spin cycles contribute to the superpotential once one
considers all solutions to~\eqref{eq:invariance-condition}.

Finally, we expect the non-perturbative effects discussed in here to
play an important role in the stabilization of K\"ahler moduli in type
IIB compactifications. As an example, in \cite{Berglund:2012xx} we
show how the $O(1)$ instanton discussed in this paper can be used to
stabilize the K\"ahler moduli in a model introduced
in~\cite{Balasubramanian:2012wd}.

\acknowledgments

We thank A.~Collinucci, R.~Donagi, R.~Savelli and A.~Uranga for
illuminating discussions. P.B. thanks the hospitality of the Simons
Center for Geometry, the Niels Bohr Institute, and the theory groups
at CERN and DESY, where much of this work was carried
out. I.G.-E. would like to thank N.~Hasegawa for kind encouragement
and support. The work of P.B. is supported by NSF grants PHY-0645686
and PHY-1207895.

\bibliographystyle{JHEP}
\bibliography{refs}

\providecommand{\href}[2]{#2}\begingroup\raggedright\begin{thebibliography}{10}

\bibitem{Denef:2004dm}
F.~Denef, M.~R. Douglas, and B.~Florea, {\it {Building a better racetrack}},
  {\em JHEP} {\bf 0406} (2004) 034,
  [\href{http://xxx.lanl.gov/abs/hep-th/0404257}{{\tt hep-th/0404257}}].

\bibitem{Balasubramanian:2005zx}
V.~Balasubramanian, P.~Berglund, J.~P. Conlon, and F.~Quevedo, {\it
  {Systematics of moduli stabilisation in Calabi-Yau flux compactifications}},
  {\em JHEP} {\bf 0503} (2005) 007,
  [\href{http://xxx.lanl.gov/abs/hep-th/0502058}{{\tt hep-th/0502058}}].

\bibitem{Freed:1999vc}
D.~S. Freed and E.~Witten, {\it {Anomalies in string theory with D-branes}},
  {\em Asian J.Math} {\bf 3} (1999) 819,
  [\href{http://xxx.lanl.gov/abs/hep-th/9907189}{{\tt hep-th/9907189}}].

\bibitem{Blumenhagen:2012kz}
R.~Blumenhagen, X.~Gao, T.~Rahn, and P.~Shukla, {\it {A Note on Poly-Instanton
  Effects in Type IIB Orientifolds on Calabi-Yau Threefolds}},  {\em JHEP} {\bf
  1206} (2012) 162, [\href{http://xxx.lanl.gov/abs/1205.2485}{{\tt
  arXiv:1205.2485}}].

\bibitem{Cicoli:2012vw}
M.~Cicoli, S.~Krippendorf, C.~Mayrhofer, F.~Quevedo, and R.~Valandro, {\it
  {D-Branes at del Pezzo Singularities: Global Embedding and Moduli
  Stabilisation}},  \href{http://xxx.lanl.gov/abs/1206.5237}{{\tt
  arXiv:1206.5237}}.

\bibitem{Sharpe:2003dr}
E.~Sharpe, {\it {Lectures on D-branes and sheaves}},
  \href{http://xxx.lanl.gov/abs/hep-th/0307245}{{\tt hep-th/0307245}}.

\bibitem{Katz:2002gh}
S.~H. Katz and E.~Sharpe, {\it {D-branes, open string vertex operators, and Ext
  groups}},  {\em Adv.Theor.Math.Phys.} {\bf 6} (2003) 979--1030,
  [\href{http://xxx.lanl.gov/abs/hep-th/0208104}{{\tt hep-th/0208104}}].

\bibitem{Diaconescu:2006id}
D.-E. Diaconescu, A.~Garcia-Raboso, R.~L. Karp, and K.~Sinha, {\it {D-Brane
  Superpotentials in Calabi-Yau Orientifolds}},  {\em Adv.Theor.Math.Phys.}
  {\bf 11} (2007) 471--516, [\href{http://xxx.lanl.gov/abs/hep-th/0606180}{{\tt
  hep-th/0606180}}].

\bibitem{Blumenhagen:2009qh}
R.~Blumenhagen, M.~Cveti\v{c}, S.~Kachru, and T.~Weigand, {\it {D-Brane
  Instantons in Type II Orientifolds}},  {\em Ann.Rev.Nucl.Part.Sci.} {\bf 59}
  (2009) 269--296, [\href{http://xxx.lanl.gov/abs/0902.3251}{{\tt
  arXiv:0902.3251}}].

\bibitem{Aspinwall:2004jr}
P.~S. Aspinwall, {\it {D-branes on Calabi-Yau manifolds}},
  \href{http://xxx.lanl.gov/abs/hep-th/0403166}{{\tt hep-th/0403166}}.

\bibitem{Cvetic:2009ah}
M.~Cveti\v{c}, I.~Garc\'ia-Etxebarria, and R.~Richter, {\it {Branes and
  instantons at angles and the F-theory lift of O(1) instantons}},  {\em AIP
  Conf.Proc.} {\bf 1200} (2010) 246--260,
  [\href{http://xxx.lanl.gov/abs/0911.0012}{{\tt arXiv:0911.0012}}].

\bibitem{KatzFiniteness}
S.~Katz, {\it On the finiteness of rational curves on quintic threefolds},
  {\em Compositio Mathematica} {\bf 60} (1986), no.~2 151--162.

\bibitem{Ganor:1996pe}
O.~J. Ganor, {\it {A Note on zeros of superpotentials in F theory}},  {\em
  Nucl.Phys.} {\bf B499} (1997) 55--66,
  [\href{http://xxx.lanl.gov/abs/hep-th/9612077}{{\tt hep-th/9612077}}].

\bibitem{Franco:2007ii}
S.~Franco, A.~Hanany, D.~Krefl, J.~Park, A.~M. Uranga, {\em et.~al.}, {\it
  {Dimers and orientifolds}},  {\em JHEP} {\bf 0709} (2007) 075,
  [\href{http://xxx.lanl.gov/abs/0707.0298}{{\tt arXiv:0707.0298}}].

\bibitem{Aganagic:2003xq}
M.~Aganagic, K.~A. Intriligator, C.~Vafa, and N.~P. Warner, {\it {The glueball
  superpotential}},  {\em Adv. Theor. Math. Phys.} {\bf 7} (2004) 1045--1101,
  [\href{http://xxx.lanl.gov/abs/hep-th/0304271}{{\tt hep-th/0304271}}].

\bibitem{Intriligator:2003xs}
K.~A. Intriligator, P.~Kraus, A.~V. Ryzhov, M.~Shigemori, and C.~Vafa, {\it {On
  low rank classical groups in string theory, gauge theory and matrix models}},
   {\em Nucl. Phys.} {\bf B682} (2004) 45--82,
  [\href{http://xxx.lanl.gov/abs/hep-th/0311181}{{\tt hep-th/0311181}}].

\bibitem{Aganagic:2007py}
M.~Aganagic, C.~Beem, and S.~Kachru, {\it {Geometric Transitions and Dynamical
  SUSY Breaking}},  {\em Nucl. Phys.} {\bf B796} (2008) 1--24,
  [\href{http://xxx.lanl.gov/abs/0709.4277}{{\tt arXiv:0709.4277}}].

\bibitem{GarciaEtxebarria:2008iw}
I.~Garc\'ia-Etxebarria, {\it {D-brane instantons and matrix models}},  {\em
  JHEP} {\bf 07} (2009) 017, [\href{http://xxx.lanl.gov/abs/0810.1482}{{\tt
  arXiv:0810.1482}}].

\bibitem{Bianchi:2011qh}
M.~Bianchi, A.~Collinucci, and L.~Martucci, {\it {Magnetized E3-brane
  instantons in F-theory}},  {\em JHEP} {\bf 1112} (2011) 045,
  [\href{http://xxx.lanl.gov/abs/1107.3732}{{\tt arXiv:1107.3732}}].

\bibitem{Berglund:2012xx}
P.~Berglund and I.~Garc\'ia-Etxebarria, {\it {K\"ahler Moduli Stabilization in
  Calabi-Yau Manifolds with Branes at Toric Singularities}}, . to appear.

\bibitem{Balasubramanian:2012wd}
V.~Balasubramanian, P.~Berglund, V.~Braun, and I.~Garcia-Etxebarria, {\it
  {Global embeddings for branes at toric singularities}},
  \href{http://xxx.lanl.gov/abs/1201.5379}{{\tt arXiv:1201.5379}}.

\end{thebibliography}\endgroup

\end{document}